\documentclass{aastex631}

\newcommand{\xb}[1]{\textsubscript{#1}}
\shorttitle{AASTeX v6.31 Sample article}
\shortauthors{Gao et al.}
\usepackage{booktabs}
\usepackage{multirow}

\begin{document}

%\title{The Energy-dependent Frequency Broadening in the Fourier Power Spectra for the X-ray Variability of MAXI J1820++070}

\title{On the Broadening of the Characteristic Frequency Range towards Higher Photon Energies in the X-ray Variability of the Black Hole Transient MAXI J1820$+$070}

%\title{Peculiar energy-dependence of power spectral components in the X-ray variability of the black hole transient MAXI J1820++070}

\correspondingauthor{Wenfei Yu}
\email{wenfei@shao.ac.cn}

\author[0000-0002-0786-7307]{ChenXu Gao}
\affiliation{Shanghai Astronomical Observatory, Chinese Academy of Sciences, \\ 80 Nandan Road, Shanghai 200030, China }
\affiliation{University of Chinese Academy of Sciences, 19A Yuquan Road, Beijing 100049, China}

\author[0000-0002-3844-9677]{Wenfei Yu}
\affiliation{Shanghai Astronomical Observatory, Chinese Academy of Sciences, \\ 80 Nandan Road, Shanghai 200030, China }

\author[0000-0002-5385-9586]{Zhen Yan}
\affiliation{Shanghai Astronomical Observatory, Chinese Academy of Sciences, \\ 80 Nandan Road, Shanghai 200030, China }

\begin{abstract}
Energy-dependent X-ray power spectral states and Band-Limited Noise (BLN) components have been seen in the low-hard state and intermediate states of black hole X-ray binaries. Here we report our analysis of \emph{Insight}-HXMT observations of the black hole transient MAXI J1820$+$070 during its 2018 outburst when the source was brightest. We found opposite trends of low-frequency ($<$ 0.1 Hz) and high-frequency ($>$ 10 Hz) BLN components, i.e., decreasing vs. increasing in frequency with increasing photon energy, respectively. This establishes an apparent two-way broadening of the power spectral plateau formed by multiple BLNs towards higher photon energies. The trend of the highest BLN component with increasing photon energy has been interpreted as that the corresponding seed photons originated from a region relatively more central in the corona previously. The decreasing trend of the characteristic frequency of the lowest frequency BLN component with increasing photon energy can then be interpreted as that the corresponding seed photons originated from further out in the disk but on the opposite side of the central corona to the observer. These opposite trends then imply that the power spectral plateau represents the radial extension of the accretion disk that contributed the seed photons producing the BLNs, and show that the higher the photon energy is, the wider the plateau and the smaller the fractional variability. The plateau shows the analogy to the flat power spectrum with a low fractional variability of the Power-Law Noise seen in the high-soft state, which corresponds to photons from the entire x-ray disk.

\end{abstract}

%\begin{abstract}
%Energy-dependence of the characteristic frequencies of the X-ray power spectral components and the power spectral states have been seen in the LHS and the intermediate state of black hole X-ray binaries. Here we report our analysis of the \emph{Insight}-HXMT observations of the black hole transient MAXI J1820++070 during its 2018 outburst when its hard X-ray intensity was among the highest in the entire outburst. We found distinct trends, i.e., decreasing vs. increasing, in the evolution of the Band-Limited Noise (BLN) components at the lowest and at the highest frequencies with photon energy up to beyond 200 keV, respectively. On the othr hand, the fractional {\it rms} of the two BLNs decreases with photon energy. While the increasing trend of the characteristic frequencies of the highest BLN component with energy can be interpreted as due to a corona with density or temperature gradient from which the higher energy photons originate from seed photons in regions more closer to the central black hole, the lowest BLN can be explained as photons from an extended disk corona further out.  The distinct trend mimics We discuss the implications for the picture of energy-dependent power spectral states. 
%\end{abstract}


\section{Introduction} \label{sec:intro} 
Most known black hole X-ray binaries (BH XRBs) in our galaxy are transient low-mass X-ray binaries (LMXBs). They spent most of their time in quiescence and occasionally turned into an accretion outburst probably due to disk instability. During such an outburst, the corresponding X-ray spectral and timing properties usually evolve significantly on timescales from days to months \citep[see reviews by ][]{remillard_x-ray_2006,done_modelling_2007,bambi_transient_2016}. Different X-ray spectral states have been classified based on the X-ray spectral and timing properties seen with X-ray observations \citep{homan_correlated_2001}. Among them, the two main spectral states are the low-hard state (LHS) and high-soft state (HSS) \citep{tananbaum_observation_1972}.

Strong X-ray variability in black hole X-ray binaries has been detected in a wide range of time scales during X-ray binary outbursts when the X-ray intensity is high enough, whereas the shortest time scales of the variability seen in the X-ray band is comparable to the dynamical time scales in the vicinity of the compact objects\citep{van_der_klis_rapid_2006}. The conventional way for examining the underlying variability components of BHXB fluctuations is the analysis through the Power Density Spectrum (PDS), which enables the investigation of both quasi-periodic oscillations (QPOs) and broadband noise in the Fourier frequency domain. The power spectrum in black hole HSS is characterized primarily by a Power-law Noise (PLN), while the power spectrum in black hole LHS is primarily dominated by one or more Band-Limited Noise (BLN) components with additional QPOs. The PDS corresponding to the intermediate BH state consists of BLNs and QPOs, but sometimes a weak PLN component also shows\citep{belloni_unified_2002}. In BH XRBs, low-frequency QPOs (LFQPOs) are prominent components in the power spectra. They are divided into three types. Type-A QPOs are relatively weak and rare, predominantly observed during the HSS; type-B QPOs are stronger and more common, occurring in the soft-intermediate state and often accompanied by PLN; Type-C QPOs, detected in the hard and hard-intermediate states, are characterized by simultaneous BLN component(s) and are frequently displaying harmonics and sub-harmonics \citep{casella_study_2004, bambi_transient_2016}.

The popular picture of correlated spectral and temporal states is significantly challenged by the observations of the black hole X-ray binary candidate MAXI J1659$-$152 \citep{yu_energy-dependent_2013}. In that study, the authors have found pronounced energy-dependent characteristics in its PDS. Specifically, during the transition from the hard state to the intermediate state, the PDS in the soft X-ray energy band below 2~ keV, which corresponds that of the disk spectral component, exhibited a PLN that is typical of the HSS, while the PDS in the hard X-ray energy band ($>$ 2 keV), corresponding to the Comptonized component, exhibited BLNs and QPOs, typical characteristics of the LHS. This demonstrates that black hole power spectral states depend actually on which spectral components we are looking at. In other words, there is no actual correspondence between the energy spectral state and the power spectral state. Moreover, subsequent studies on the energy dependence of the PDSs of GRS 1915+105 yielded similar findings \citep{stiele_detection_2014}. These studies further suggest that to trace the emergence of the thermal disk component in the soft X-ray energy band ($<$ 2 keV), timing analysis is more sensitive than spectral analysis.

Through power spectral analysis, \cite{wijnands_broadband_1999} discovered a strong correlation between the QPO frequency (the low-frequency QPO) and the break frequency of BLN in various types of X-ray binaries. This suggests that BLN and QPO may originate from the same physical mechanism or somehow couple in frequency under a certain unknown mechanism. On the other hand, \cite{yu_hard_2003} show that the spectral transitions in the neutron star LMXB transient Aquila X-1 share a great similarity to those of black hole transients; they found that the primary BLN and the low-frequency QPO in both black hole and neutron star soft X-ray transients follow an identical correlation between the break frequency of the BLN and the frequency of the low-frequency QPO. This demonstrates that the BLN and the QPO in the BH X-ray binaries are of the same origins as those in the neutron star systems \citep{yu_hard_2003}. Correlation between the kHz QPO frequency and the X-ray count rate on the BLN time scales in the Z-type neutron star LMXB Sco X-1 have demonstrated that the BLN corresponds to the modulation in the mass accretion rate. In addition, in the atoll type neutron star low-mass X-ray binaries, the frequencies of the kilohertz QPOs correlate with X-ray intensity on time scales from seconds to a few tens of seconds, i.e., on the BLN time scales, in observations of hours and establish the so-called 'parallel track' phenomenon \citep{yu_khz_1997, mendez_dependence_1999}. These observational studies imply that the BLN corresponds to the variation in the mass accretion rate. On the other hand, theoretical modeling of the X-ray variability of the accretion flow based on the propagation scenario has reproduced the BLN components in the PDS \citep{lyubarskii_flicker_1997, ingram_physical_2011}. Moreover, there are also notable similarities between black hole and neutron star systems from the perspectives of state transitions, energy spectral, and temporal properties\citep{gardenier_model-independent_2018, belloni_x-ray_2018}. \citet{munoz-darias_black_2014} demonstrated that all types of neutron star LMXB can also be classified according to the hard-intermediate-soft state scheme observed in BH systems. These similarities suggest that we can learn from BLNs and QPOs in individual sources or source types to achieve a universal understanding of the BLNs and QPOs in both BH and NS XRBs.

\citet{stiele_energy_2015} has detected the energy dependence of the characteristic frequency of the BLN in GX 339$-$4 and several other black hole transients for the first time, strengthening the energy-dependent picture of the power spectral states of black hole X-ray binaries \citep{yu_energy-dependent_2013}. These results show that the characteristic frequency of the BLN increased with increasing photon energy in the energy range below 10 keV, as seen with the XMM-Newton. \citet{yang_accretion_2022} show the same energy dependence of the characteristic frequency of the BLN extends to the hard X-ray band in MAXI J1820$+$070, beyond a few tens of keV with HXMT observations. NICER observations of MAXI J1820$+$070 \citep{kawamura_full_2022} have also shown the energy dependence of the characteristic frequency of the BLN at the highest frequency, confirming the energy dependence of the BLN in BH XRBs previously found \citep{stiele_detection_2014}. In \cite{gao_low_2023}, we analyzed the modulation of spectral components in different phases of the LFQPO during a hard-state observation of MAXI J1820$+$070. %We suggested that the original beats of the LFQPO come from the Compton component. 
In the above analysis, we detected the energy dependence of the BLN at the highest frequency in MAXI J1820$+$070 with \emph{Insight}-HXMT that was reported in \cite{yang_accretion_2022}, but with additional discoveries regarding the behavior of the BLN component at the low frequency end. Here we present our findings of the energy dependence of the BLNs in a broader frequency range with the \emph{Insight}-HXMT observations.

% 1820 的介绍

\section{Observations and Data reduction} \label{sec:analysis}

The black hole transient MAXI J1820$+$070 was first discovered in X-ray emissions on 11 March 2018, by the Monitor of All-sky X-ray Image \citep[MAXI; ][]{kawamuro_maxigsc_2018,tucker_asassn-18ey_2018}. During the onset of the outburst, its luminosity exhibited a rapid rise, reaching a level of 2 Crab within the energy range of 2-20 keV \citep[][]{shidatsu_x-ray_2018}. Subsequently, the source sustained a prolonged period of approximately three months in the black hole LHS, characterized by a slight decline in luminosity \citep{shidatsu_x-ray_2019}. The long-lived LHS provides an exceptional opportunity to investigate its X-ray timing and spectral behavior \citep[e.g.][]{buisson_maxi_2019,kara_corona_2019}, especially in the hard X-ray energies with \emph{Insight}-HXMT \citep{ma_discovery_2020}. 

\subsection{Observations}
\emph{Insight}-HXMT consists of three payloads: high-energy X-ray telescope (HE, 35-250 keV), medium-energy X-ray telescope (ME, 10-35 keV), and low-energy X-ray telescope (LE, 1-10 keV)\citep{zhang_overview_2020}.
The technical details of \emph{Insight}-HXMT and data analysis guide can be seen in the data reduction guide\footnote{http://hxmten.ihep.ac.cn/SoftDoc/501.jhtml}. In our analysis, we utilized the \emph{Insight}-HXMT Data Analysis Software (HXMTDAS) version 2.05, specifically designed for \emph{Insight}-HXMT data processing. To ensure minimize influence from neighboring sources, we opted for narrow field of views (FOVs) for all three telescopes. Furthermore, we followed the recommended criteria outlined in the guide for producing accurate and reliable Good-Time Intervals (GTIs).

\subsection{Data reduction}
In the analysis, we computed the Miyamoto normalized PDSs for light curves in different energy bands\citep{miyamoto_large_1995}. For data obtained with the LE and ME instruments, we utilized a time interval of 512 seconds and a time resolution of 1/512 seconds, resulting in a Nyquist frequency of 256 Hz. For data obtained with the HE instruments, we employed a time interval of 512 seconds and a time resolution of 1/2048 seconds, corresponding to a Nyquist frequency of 1048 Hz.

For the PDS corresponding to the LE and ME instruments, we calculated the average power above 64 Hz in the PDS and took it as the white noise level. For the HE instrument, we specifically generated the PDS of the photon series obtained with the blocked collimator to ensure an accurate determination of the white noise level, free from potential background model biases. We then calculated the average power above 64 Hz in this PDS, which was taken as the white noise level for the PDS in the HE instrument.

\begin{figure}[ht!]
    \plotone{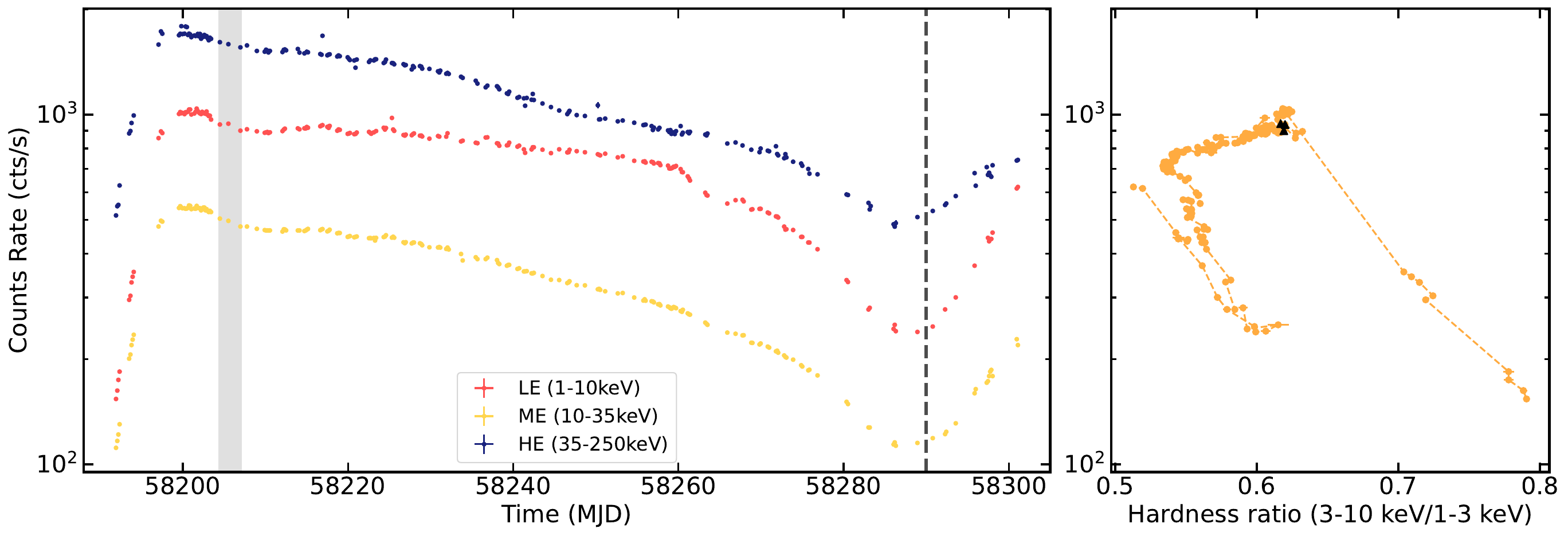}
    \caption{Light curves (left) and HID (right) traced by MAXI J1820$+$070 during its LHS in the 2018 outburst based on observations of \emph{Insight}-HXMT. In the left panel, the black dashed line marks the end of the conventional X-ray LHS. The observations used in this analysis are highlighted with a shaded region in the left panel and marked with black triangles in the right panel. In the right panel, the vertical axis represents the 1-10 keV count rate.
    \label{fig:lc_HID}}
\end{figure}

\begin{deluxetable*}{lcc}
\tablenum{1}
\caption{\emph{Insight}-HXMT data information used in this analysis.
    \label{tab:info}}
\tablewidth{0pt}
\tablehead{
\colhead{Exposure ID} & \colhead{Start Time (UTC)} & \colhead{Exposure (s)}
}
\decimalcolnumbers
\startdata
P011466100601 & 2018/03/27 08:29:37 & 3325 \\
P011466100701 & 2018/03/28 09:56:48 & 2550 \\
P011466100801 & 2018/03/29 20:56:55 & 4889 \\
\enddata
\end{deluxetable*}

\section{Power spectral analysis and Results}
\begin{figure}[ht!]
    \plotone{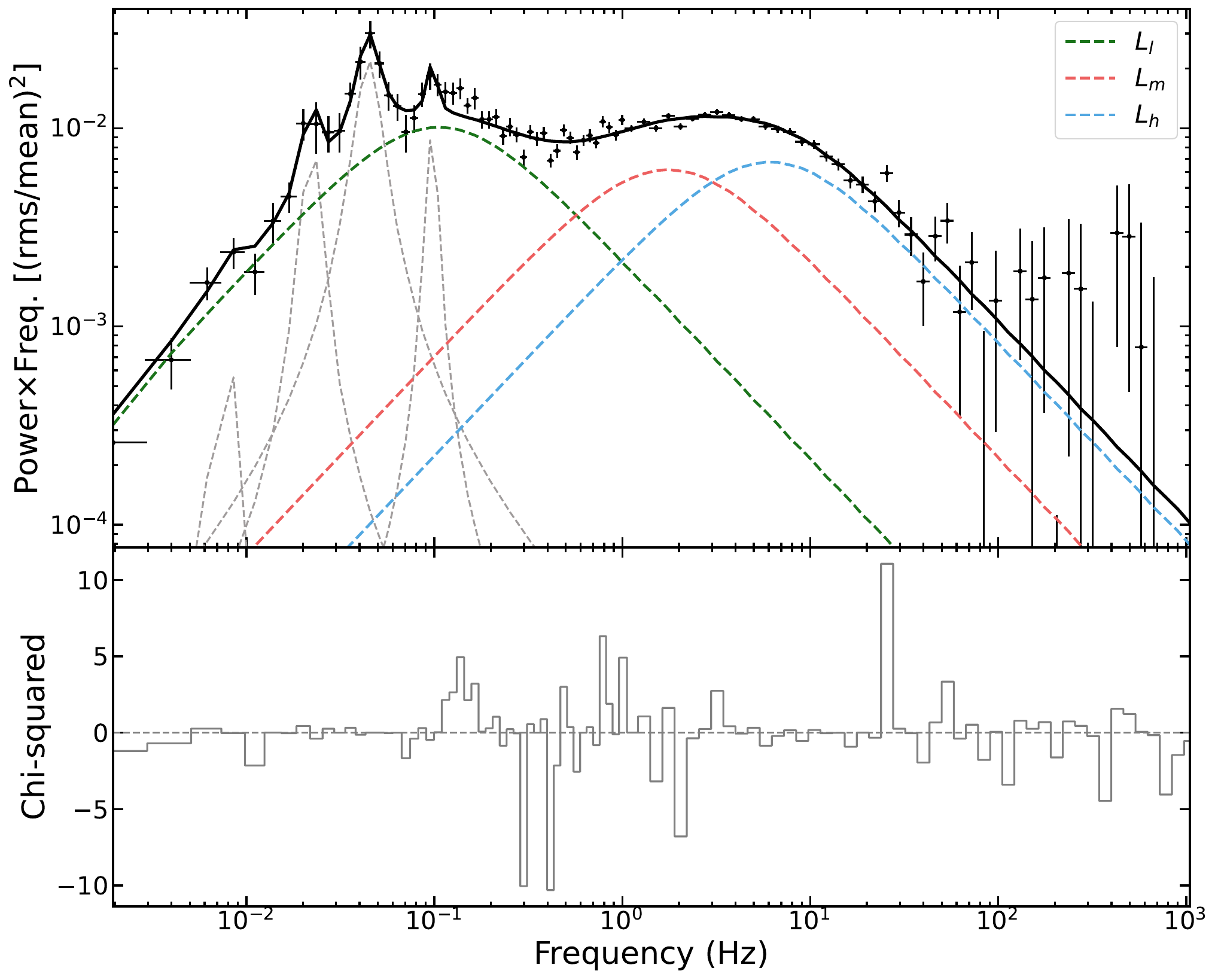}
    \caption{The representative PDS corresponding to the photons in the 35–46 keV band of the HE energy range from Obs-0601, with the best-fit power spectral models and individual Lorentzian components over-plotted. The light gray dashed line marks the QPO and its harmonic. The dashed lines in green, red, and blue indicate the zero-centered Lorentzian used to describe the BLN components. The bottom panel shows the residuals of the model fit. The plateau of the power spectrum is characterised by the two characteristic frequencies of the low and the high BLN components. i.e., ${\rm L}_{\it l}$ and ${\rm L}_{\it h}$.
    \label{fig:demo_pds}}
\end{figure}

We selected a series of consecutive observations near the LHS peak, characterized by the longest single exposure times and significant QPO levels, based on \emph{Insight}-HXMT data. The details of the \emph{Insight}-HXMT data used in this analysis are provided in \autoref{tab:info}. We also plotted the light curves and hardness-intensity diagram (HID) corresponding to the observations of MAXI J1820$+$070 obtained by the three \emph{Insight}-HXMT detectors (\autoref{fig:lc_HID}), with our selected time range of the observations marked in the plot. The PDSs were obtained for nine different energy bands: LE (1-2 keV, 2-5 keV, 5-10 keV), ME (10-16 keV, 16-35 keV) and HE (35-46 keV, 46-60 keV, 60-76 keV, 76-100 keV, 100-250 keV). The PDSs were rebinned in frequency for our analysis that focused on the BLNs. Since these observations were conducted during the LHS during the outburst, we chose to use three or more Lorentzians to model both the QPO and the BLN in each energy band. The PDS fitting procedure was performed using \texttt{XSPEC} version 12.11.1. In the QPO fitting, we considered both the harmonic and sub-harmonic components. To model the BLNs, we used three zero-centered Lorentzians, representing the three BLN components shown in \autoref{fig:demo_pds}: low-frequency component(L\xb{l}), medium-frequency component(L\xb{m}), and high-frequency component(L\xb{h}). Specifically, we present the results of observation P011466100601, referred to as Obs-0601, as the main representative example in this analysis. Additionally, we refer to the stack dataset of observations P011466100701 and P011466100801, denoted Obs-0708S, to corroborate and emphasize the findings from Obs-0601. This complementary dataset allows us to confirm our primary findings and explore consistencies across similar observational datasets. 

\begin{figure}[ht!]
    \plotone{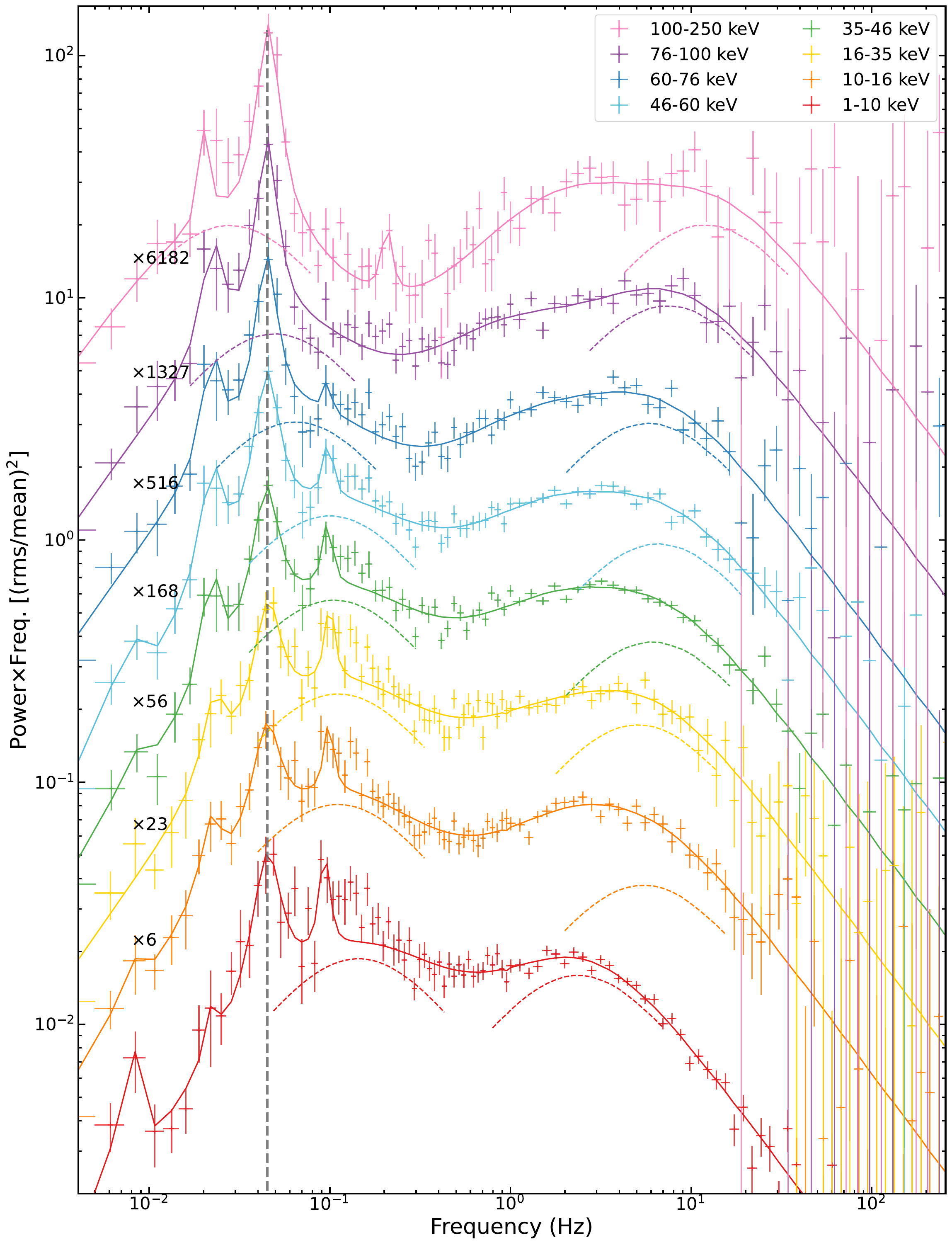}
    \caption{Simultaneous PDS corresponding to photon series in different energy bands obtained from the observation with Obs-0601. The PDS have been shifted vertically to show the trend of the characteristic frequency of the broadband noise in relative photon energy. A vertical grey dashed line marks the frequency of the BH LFQPO. Different colors are used to distinguish the data of different energy bands, with the best-fit models plotted as solid lines. The L\xb{l} and L\xb{h} are over-plotted to show their opposite trends, establishing a broader plateau of the PDS towards higher photon energies. 
    \label{fig:waterfall}}
\end{figure}

We show the characteristic frequencies and fractional root mean square (rms) values of the best-fitting results in \autoref{tab:bln} for Obs-0601. The characteristic frequencies were taken as $\nu_{\mathrm{max}} = \sqrt{\nu_0^2 + (\omega/2)^2}$, where $\nu_0$ represents the centroid frequency and $\omega$ denotes the full width at half maximum (FWHM) of the Lorentzian. We use $\nu$\xb{l}, $\nu$\xb{m}, and $\nu$\xb{h} to represent the characteristic frequencies of L\xb{l}, L\xb{m}, and L\xb{h}, respectively. Furthermore, $\nu_{\mathrm{QPO}}$ represents the frequency of the QPO.
In \autoref{fig:waterfall}, we present the power spectra illustrating the data and the best-fitted model for different energy bands of the PDS from Obs-0601. Interestingly, as we look at a higher energy band, the two prominent BLN components in the PDS display characteristic frequencies toward lower and higher frequencies, respectively. The total fractional rms of the power spectra is also significantly reduced roughly by half. To show the energy dependence of the PDS shape and the amplitude more clearly, we divided the HE energy band into 11 energy ranges. We extracted 11 additional PDS from HE with overlapping energy ranges: 40-51 keV, 45-56 keV, 50-62 keV, 55-69 keV, 60-78 keV, 65-86 keV, 70-93 keV, 75-110 keV, 80-130 keV, 85-150 keV, 90-200 keV. \autoref{fig:bln_601} shows the three characteristic frequencies of BLNs for all energy bands extracted from Obs-0601 with the same color scheme as highlighted in \autoref{fig:demo_pds}. %Similarly, \autoref{fig:stacked} presents the results for the stacked observations of Obs-0708S. 
Apparent excess power near 0.2 Hz has been visible in some of the power spectra in the low-energy bands. However, fitting with an additional Lorentzian component for this feature yields identical L\xb{l} as those in \autoref{tab:bln}, with also nearly identical uncertainties, indicating this feature has no impact our measurements of the frequency evolution.

\begin{deluxetable*}{lccccccc}
\tablenum{2}
\caption{Energy dependence of the characteristic frequency ($\nu_{\mathrm{max}}$), the fractional rms for individual BLN, and the total fractional rms for all BLNs, supplemented by Pearson correlation coefficients between the characteristic frequency and photon energies, calculated for all energy bands and for energy levels above 30 keV.
    \label{tab:bln}}
\tablewidth{0pt}
\tablehead{
\colhead{Energy band} & \multicolumn{2}{c}{L\xb{l}} & \multicolumn{2}{c}{L\xb{m}} & \multicolumn{2}{c}{L\xb{h}} & \colhead{BLNs}\\
\colhead{} & \colhead{$\nu_{\mathrm{max}}$} & \colhead{rms(\%)} & \colhead{$\nu_{\mathrm{max}}$} & \colhead{rms(\%)} &\colhead{$\nu_{\mathrm{max}}$} & \colhead{rms(\%)} & \colhead{Total rms(\%)}
}
\decimalcolnumbers
\startdata
1-2 keV (LE) & $0.12^{+0.02}_{-0.01}$ & $24.1^{+1.1}_{-1.3}$ & $1.67^{+0.11}_{-0.59}$ & $1.0^{+0.5}_{-0.5}$ & $1.81^{+0.14}_{-0.14}$ & $21.3^{+0.6}_{-0.6}$ & $32.2^{+0.9}_{-1.1}$\\
2-5 keV (LE) & $0.17^{+0.03}_{-0.02}$ & $24.4^{+1.2}_{-1.1}$ & $2.50^{+0.02}_{-0.83}$ & $1.7^{+0.2}_{-0.3}$ & $2.79^{+0.22}_{-0.17}$ & $24.1^{+0.5}_{-0.6}$ & $34.3^{+0.9}_{-0.9}$\\
5-10 keV (LE) & $0.20^{+0.04}_{-0.03}$ & $22.8^{+1.4}_{-1.4}$ & $1.81^{+0.74}_{-0.79}$ & $6.6^{+0.3}_{-0.4}$ & $3.08^{+0.51}_{-0.36}$ & $23.6^{+1.0}_{-0.8}$ & $33.5^{+1.2}_{-1.1}$\\
10-16 keV (ME) & $0.11^{+0.02}_{-0.01}$ & $20.6^{+0.8}_{-0.8}$ & $2.21^{+0.64}_{-0.40}$ & $15.4^{+0.5}_{-0.4}$ & $5.43^{+0.64}_{-1.43}$ & $14.0^{+2.4}_{-2.2}$ & $29.3^{+1.3}_{-1.2}$\\
16-35 keV (ME) & $0.11^{+0.01}_{-0.01}$ & $17.8^{+0.9}_{-0.8}$ & $1.49^{+0.53}_{-0.38}$ & $10.9^{+0.5}_{-0.5}$ & $5.06^{+0.62}_{-0.87}$ & $15.4^{+1.2}_{-1.4}$ & $25.9^{+1.0}_{-1.0}$\\
35-46 keV (HE) & $0.11^{+0.01}_{-0.01}$ & $17.8^{+0.7}_{-0.7}$ & $1.73^{+0.29}_{-0.28}$ & $13.9^{+0.2}_{-0.4}$ & $6.08^{+0.91}_{-0.79}$ & $14.6^{+1.2}_{-1.2}$ & $26.9^{+0.8}_{-0.8}$\\
46-60 keV (HE) & $0.10^{+0.02}_{-0.01}$ & $15.3^{+0.8}_{-0.7}$ & $1.69^{+0.27}_{-0.49}$ & $13.0^{+0.3}_{-1.6}$ & $6.55^{+1.14}_{-1.79}$ & $13.5^{+1.9}_{-0.2}$ & $24.2^{+1.2}_{-1.0}$\\
60-76 keV (HE) & $0.07^{+0.01}_{-0.01}$ & $13.7^{+0.7}_{-0.9}$ & $1.33^{+0.39}_{-0.32}$ & $10.8^{+0.2}_{-0.5}$ & $5.70^{+0.06}_{-1.20}$ & $13.6^{+1.1}_{-0.9}$ & $22.1^{+0.8}_{-0.8}$\\
76-100 keV (HE) & $0.05^{+0.01}_{-0.01}$ & $13.0^{+0.7}_{-1.0}$ & $1.02^{+0.05}_{-0.42}$ & $11.1^{+0.4}_{-0.5}$ & $7.46^{+2.24}_{-2.62}$ & $14.9^{+0.9}_{-0.6}$ & $22.7^{+0.7}_{-0.7}$\\
100-250 keV (HE) & $0.03^{+0.01}_{-0.00}$ & $10.1^{+0.7}_{-0.8}$ & $2.16^{+0.61}_{-0.52}$ & $10.5^{+0.3}_{-0.4}$ & $11.93^{+6.58}_{-4.62}$ & $10.0^{+1.6}_{-1.5}$ & $17.7^{+1.0}_{-1.0}$\\
\hline
Pearson & \(-0.94\) & -- & \(0.53\) & -- & \(0.90\) & -- & --\\
Pearson ($>$ 30 keV) & \(-0.89\) & -- & \(-0.06\) & -- & \(0.79\) & -- & --\\
\enddata
\tablecomments{The main table presents characteristic frequency $\nu_{\mathrm{max}}$ and fractional rms values, including 90\% confidence intervals.}
%\tablecomments{This method approximates the sampling distribution of the correlation coefficient by transforming it into a normally distributed variable, allowing for the derivation of confidence intervals under the assumption of bivariate normality of the underlying data.}
\end{deluxetable*}

\subsection{Properties of low-frequency component L\xb{l}}
Among the three zero-centered Lorentzian components that describe the power spectrum, L\xb{l} has the lowest characteristic frequency (see \autoref{tab:bln} and \autoref{fig:bln_601}), which is around 0.1 Hz in the X-ray energy band below 30 keV. The characteristic frequency $\nu$\xb{l}, corresponding to L\xb{l}, does not show a significant trend of evolution within the X-ray energy band below 30 keV. %, consistent across the Obs-0708S(see \autoref{fig:stacked}).  Across all energy bands, the QPO frequency does not change and is around 0.044 Hz. 
Compared with the QPO frequency measured simultaneously, the $\nu$\xb{l} in the X-ray energy band below 30 keV is slightly more than double the QPO frequency. Starting from 30~keV, $\nu$\xb{l} begins to decrease with increasing energy. Between 60~keV and 70~keV, $\nu$\xb{l} drops approximately twice $\nu_{\mathrm{QPO}}$, and from 90~keV or so, it gets closer to $\nu_{\mathrm{QPO}}$. In the energy band above 100 keV, $\nu$\xb{l} decreases further to around 0.03 Hz, slightly higher than half $\nu_{\mathrm{QPO}}$. This indicates a progressive decrease of L\xb{l} as compared with the QPO frequency as the energy band increases. 

The fractional rms of the L\xb{l} shows a gradual decrease with increasing energy. Specifically, the fractional rms remains relatively constant within the energy band covered by LE but starts to decrease gradually from 30~keV with increasing energy. Up to the energy band above 100 keV, the fractional rms is only half of the value observed in the 1-2 keV energy band. This suggests a decreasing fractional amplitude of the L\xb{l} with increasing energy, indicating a change in its strength or the contribution relative to the overall signal.

\subsection{Properties of high-frequency component L\xb{h} }
The L\xb{h} has the highest characteristic frequency among the three Lorentz components seen in the power spectra and is always greater than 1~Hz. $\nu$\xb{h} began to shift towards higher frequencies with increasing photon energy and started to rise from around 2~Hz in the 1-2~keV photon energy band, reaching approximately 5~Hz around 30~keV and went beyond 10~Hz beyond 100~keV. 

\begin{figure}[ht!]
    \plotone{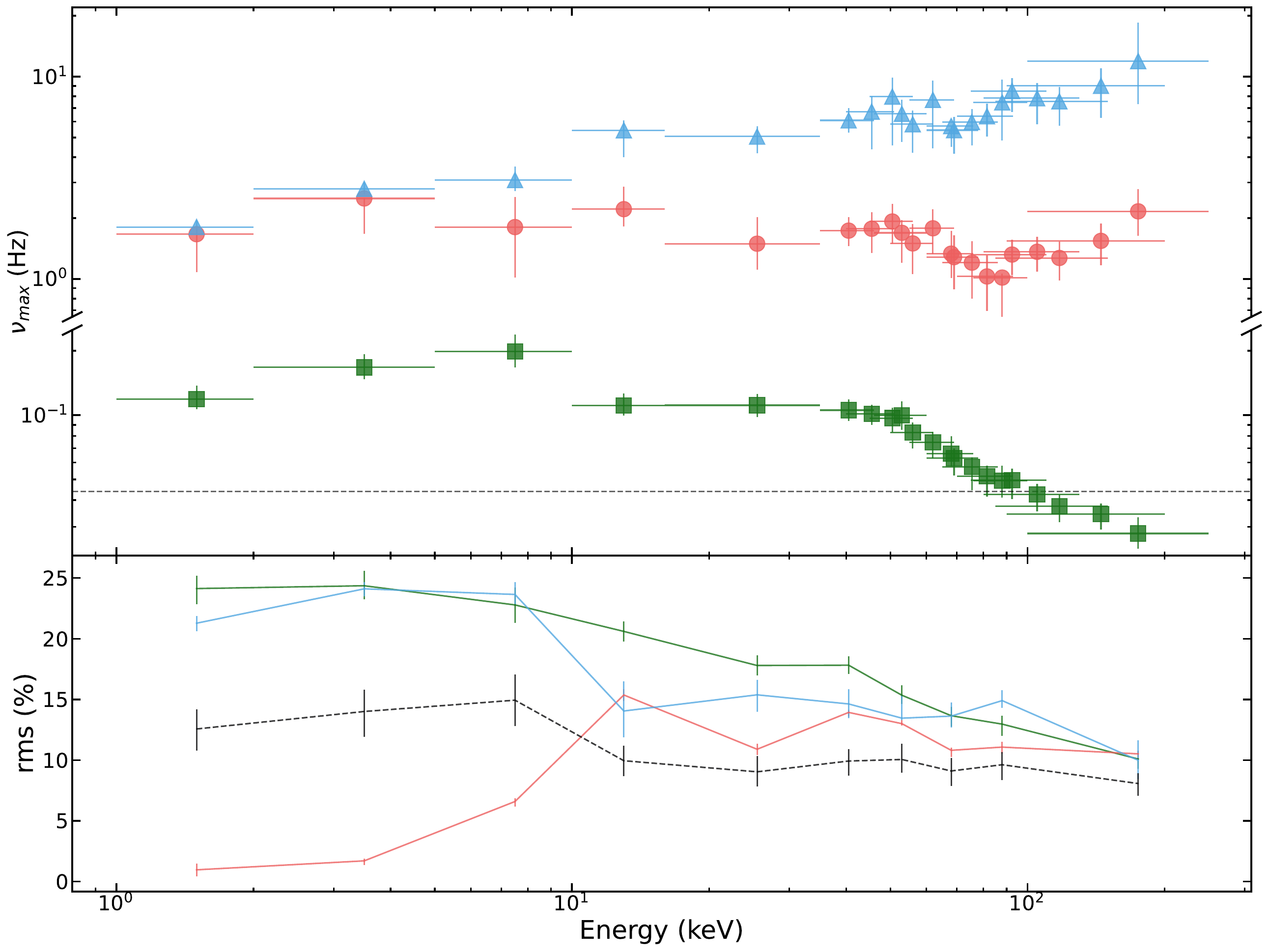}
    \caption{The energy dependence of the characteristic frequency and the fractional rms for the three BLN Lorentzian components. The color scheme of three BLN follows those highlighted in \autoref{fig:demo_pds}. The $\nu_{\mathrm{QPO}}$ is indicated by a horizontal dashed line in gray in the upper panel and its fraction rms is represented by gray dashed line in lower panel. A break in the y-axis was implemented to show the trends. Data were extracted from Obs-0601.
    \label{fig:bln_601}}   
\end{figure}

The fractional rms versus energy relation of L\xb{h} is similar to that of L\xb{l}; both did not show significant changes below 10~keV, and started to decrease by half towards higher photon energies. Furthermore, we plotted the QPO rms in the lower panel of \autoref{fig:bln_601} and found it exhibits a trend very similar to that of $\nu$\xb{h} and $\nu$\xb{l}. In \cite{gao_low_2023}, we suggest that the monotonic decrease in the trend of QPO rms above 2 keV is due to the modulation of a larger reflective component in the spectral components, with the modulating signal of the reflective component originating from the modulation of the Compton component. These results might suggest a similarity of the amplitude vs photon energy relation between the QPO and BLN signals in the Compton processes.

\subsection{Properties of medium-frequency component L\xb{m}}
The L\xb{m} falling between those of L\xb{l} and L\xb{h}. In the 1-10~keV energy band, the L\xb{m} is almost indistinguishable from L\xb{h} at around 2~Hz. However, as the frequency of the $\nu$\xb{h} shifts towards higher frequencies with increasing energy, $\nu$\xb{m} remains relatively stable in frequency; the two noise components gradually differ in frequency towards higher photon energies.

Since L\xb{l} and L\xb{h} shift towards lower and higher frequencies, respectively, L\xb{m} contributes to how flat the plateau between the two extreme components is. In the 1-10~keV energy band, the L\xb{m} has a very small fractional rms ($\lesssim$ 5\%). As the photon energy increases, L\xb{m} becomes more pronounced; its fractional rms in the energy band above 35~keV is comparable to that of L\xb{l} and L\xb{h}.

\subsection{Further Analysis}
The opposite trends in the evolution of the characteristic frequencies $\nu$\xb{l} and $\nu$\xb{h} with photon energy are more obvious in our plot of the ratio of $\nu$\xb{h} to $\nu$\xb{l} as a function of energy (see \autoref{fig:freq_ratio}). The results reveal that the ratio remains below 20 for photon energies below $\sim$20 keV but increases significantly to 400 when the photon energy rises from 10 keV to above 100 keV. The frequency range covered by the BLN components broadens in both directions as photon energy increases, exhibiting a wider plateau in the range between $2\times{10}^{-2}$ Hz and 10 Hz for photon energies beyond $\sim$30 keV. We also fitted the PDSs which cover the frequency range as low as 1/512 Hz by a model composed of four nonzero centered Lorentzian components, which was applied in previous studies \citep{yang_accretion_2022}. We still found the characteristic frequency of the lowest Lorentzian component decreases with increasing energy, from $0.048^{+0.010}_{-0.011}$ Hz in the 1--10 keV band to $0.025^{+0.002}_{-0.002}$ Hz in the 10--76 keV band and $0.023^{+0.004}_{-0.003}$ Hz in the 76--250 keV band. This demonstrates that the energy-dependent trend is independent of the PDS model used.

\begin{figure}[ht!]
    \plotone{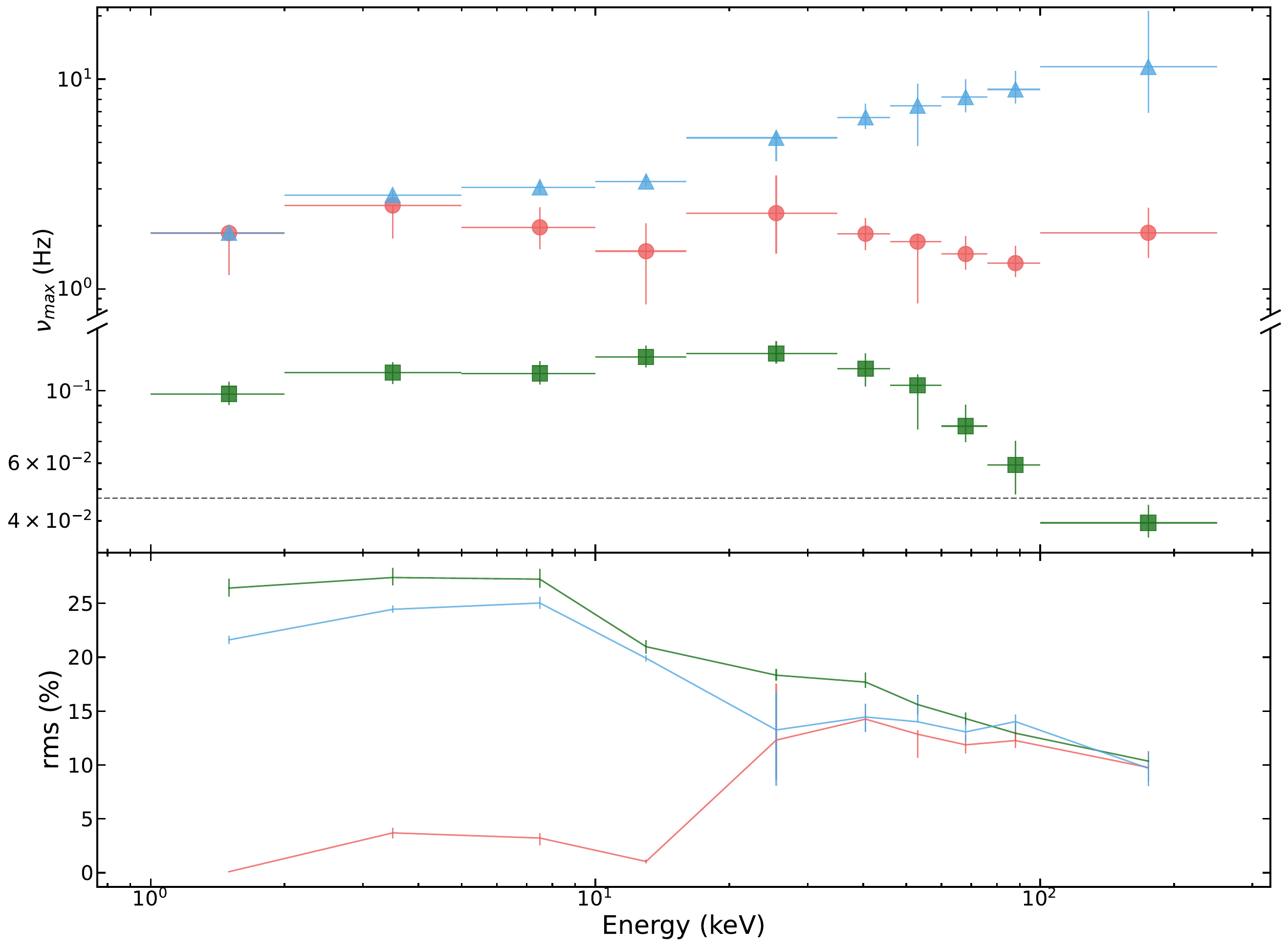}
    \caption{The energy dependence of the characteristic frequencies and fractional rms for the three BLN Lorentzian components is shown, with the colors corresponding to those highlighted in \autoref{fig:demo_pds}. The $\nu_{\mathrm{QPO}}$ is indicated by a horizontal dashed line in gray. Data in the plot were derived from the average PDS of the stacked observations Obs-0708S. 
    \label{fig:stacked}}
\end{figure}

\begin{figure}[ht!]
    \plotone{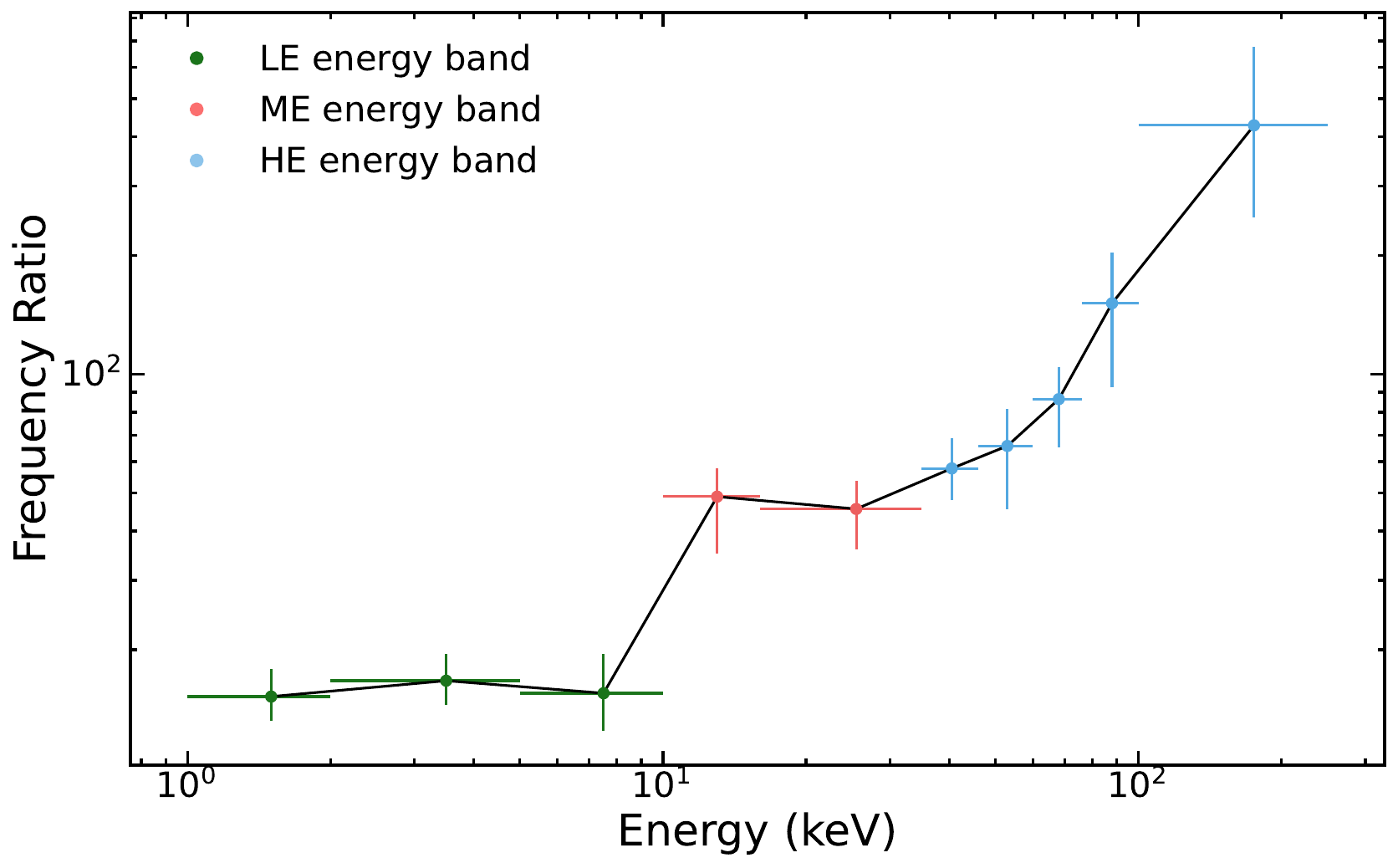}
    \caption{Ratio between the characteristic frequency of the L\xb{h} to that of the L\xb{l} components, indicating a significant broadening of the flat plateau of broadband noise in the PDS. Different colors represent data from different instruments.}
    \label{fig:freq_ratio}
\end{figure}

The Pearson correlation coefficients for the characteristic frequencies of each BLN component as a function of photon energy are presented in \autoref{tab:bln}, which show a strong negative correlation between $\nu$\xb{l} and photon energy, a strong positive correlation for $\nu$\xb{h}, and a insignificant correlation for $\nu$\xb{m}. These findings highlight the energy dependence of the frequency range of the BLN plateau and its total fractional rms. We also found the observation of Obs-0708S showed similar results as what we found above. The detailed results are shown in \autoref{fig:stacked}. This includes the same quantitative energy dependence exhibited by the BLN characteristic frequencies as well as the fractional rms spectra. 

%For enegies above 30 keV, the correlations for $\nu$\xb{m} is reduced, while those for $\nu$\xb{l} and $\nu$\xb{h} maintain notable correlations, albeit reduced in strength. These results confirm the marked energy dependence of the BLN components, particularly in the higher energy ranges ($>$30 keV). 

%Specifically, the Obs-0708S data also show that $\nu$\xb{l} decreases with increasing photon energy, $\nu$\xb{h} increases with increasing photon energy, while $\nu$\xb{m} remains stable within the 1--200 keV range. In addition, the trend of fractional RMS of L\xb{l} and L\xb{m} decreasing with increasing photon energy is also confirmed in the Obs-0708S data, where the fractional RMS decreases from about 25\% in the low-energy range to around 10\% in the high-energy range, consistent with the results from Obs-0601. This agreement between the two observations is consistent with out findings.

\subsection{Summary}
We have analyzed the energy dependence of the BLN components in the X-ray power spectra of the black hole transient MAXI J1820$+$070 in the energy range of 1--250 keV obtained with \emph{Insight}-HXMT observations. We found a two-way broadening of the plateau established by the lowest and highest BLN components in the X-ray power spectra. Specifically, the characteristic frequency $\nu$\xb{l} decreases with increasing photon energy in the range above 30~keV, dropping from 0.1~Hz at 1--2~keV to below 0.03~Hz at 100--250~keV; $\nu$\xb{m} remains relatively stable around 1--2~Hz, almost independent of photon energy; and $\nu$\xb{h} increases from around 1~Hz at the 1--2 keV band to approximately 10~Hz in the energy band beyond 100~keV. With increasing photon energy, the low frequency shoulder of the BLN plateau reduced by a factor of 3 or more in frequency, while the high frequency shoulder increase by a factor of 5 or more. As indicated in \autoref{fig:waterfall} and \autoref{tab:bln}, the total fractional rms is also reduced with increasing photon energy, from above 30\% in 1-10 keV to about 18\% in the energy range above 100 keV. Therefore, accompanied with the broadening in the frequency range of the BLN plateau by more than a decade, the total fractional rms of the BLNs reduced by half, to less than 18\% in the energy band above 100~keV.

% figure 3

\begin{figure*}
\centering
\includegraphics[angle=-90,width=1.0\textwidth]{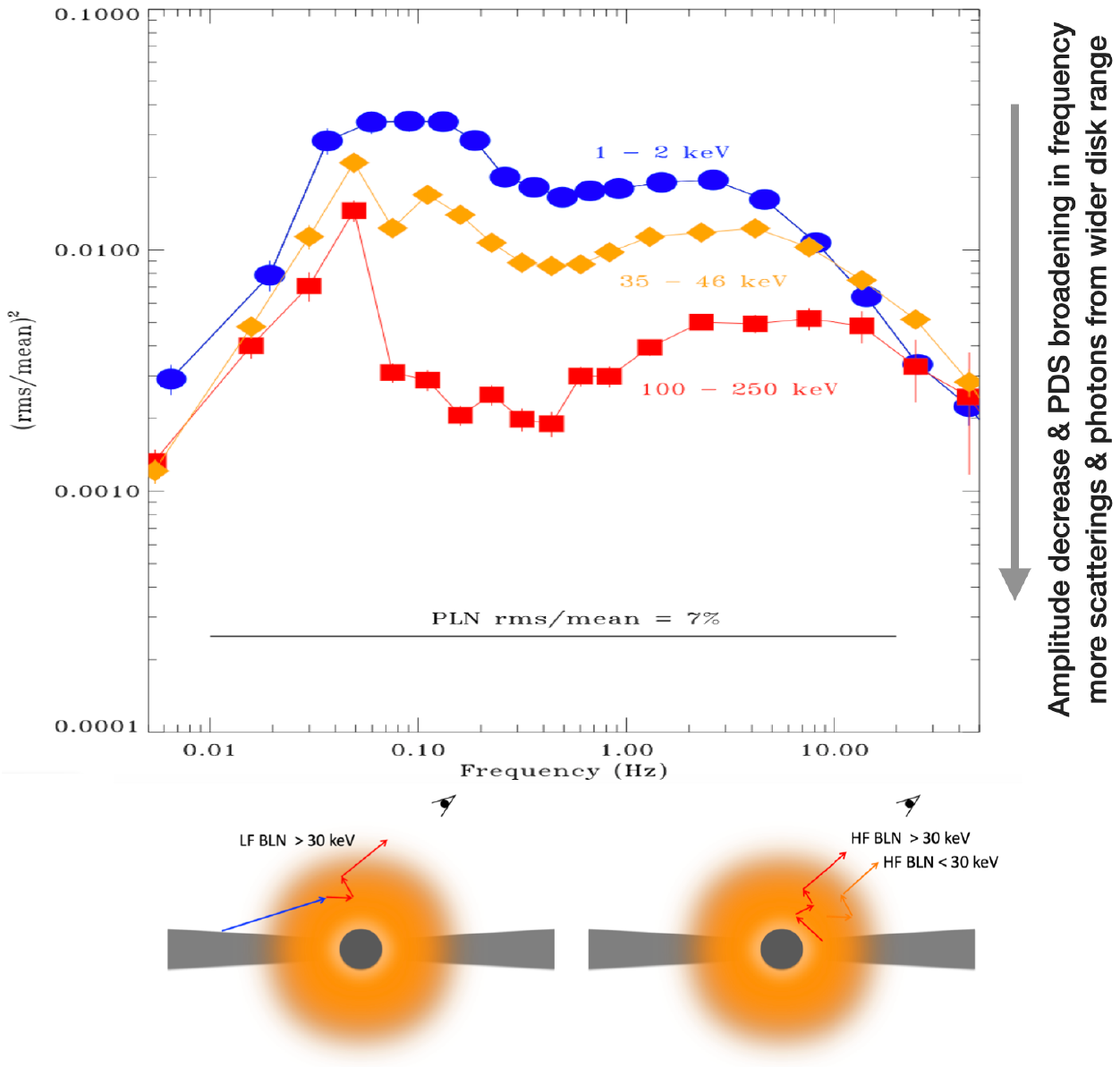}
    \caption{The schematic picture of the correspondence between the Comptonization geometries (lower) and the observed energy-dependent power spectra (upper panel) extracted in 1$-$2 keV (blue), 35$-$46 keV (orange), and 100$-$250 keV (red) from Obs-0601. The characteristic frequency of either the high-frequency BLN component ${\rm L}_{h}$ or the low-frequency BLN component ${\rm L}_{l}$ depends on the photon energy, which traces the radii at which the seed photons originate. A representative PLN with a fractional rms of 7\% is shown as a horizontal solid line, indicating the PDS of BH HSS with approximately the maximal variability observed \citep{mao_power_2021}.  
    \label{fig:interpretation}}
\end{figure*}

% Yu&zhang,2013 和 stiele&yu,2014
% 物理图像
\section{Discussion and Conclusion}

The black hole power spectral states have been found dependent on the specific spectral components we are looking at \citep{yu_energy-dependent_2013}. Specifically, MAXI J1659-152 showed the most extreme transition of the characteristic frequency of the X-ray variability when the photon energy crosses the boundary $\sim$ 2~keV between the disc component and the hard power-law spectral component \citep{yu_energy-dependent_2013}; the PDS transited from a PLN usually seen in the HSS to a BLN plus QPO type usually seen in LHS or intermediate state. Similar energy dependence of the power spectral states has also been seen in GRS 1915+105 \citep{stiele_detection_2014}. It is worth noting that a PLN can be interpreted as either the characteristic variability of a BLN evolves to an extremely low or extremely high frequency, or to both. Indeed, the PLN is thought to originate from photons of the X-ray emitting thin disk, from the innermost disk edge to the disk region further out, containing variability in a wider frequency range; the BLN components show variability in a relatively narrow frequency range as compared with the PLN, as those corresponding seed photons are from only the innermost disk subjective to Compton up-scattering by the corona (for example, see the schematic picture discussed in \citet{stiele_energy_2015}). The above energy dependence of the black hole power spectra reveals clues to the generation of the BLN components, as well as providing new clues to the origin of the high-energy photons in black hole binaries through Comptonization. 

\subsection{Energy-dependent Band-Limited Noise components}

Previous studies \citep{stiele_energy_2015,kawamura_full_2022,yang_accretion_2022} of L\xb{h} have shown the energy dependence of the $\nu_{h}$ in BH LHS. Our analysis show not only the same results for the high-frequency BLN, but also the opposite trend of the low-frequency BLN, the L\xb{l}, to lower frequencies with increasing photon energies. The phenomenon became significant when photon energy goes beyond around 30 keV. %It is also worth noting that its characteristic frequency has crossed the frequency of the low-frequency QPOs from above to below with increasing photon energy. In \autoref{fig:bln_601}, we illustrate the shifts in the characteristic frequency of the L\xb{l} component by showing the PDS in different energy bands. 
The 30 keV photon energy boundary roughly marks the photon energy beyond which the Compton component dominates. Since BLNs in black hole binaries and neutron star LMXBs are of the same origin \citep{wijnands_broadband_1999,yu_hard_2003}, which likely corresponds to the variation in the mass accretion rate \citep{yu_energy-dependent_2013}, the input variability signals of the BLN components likely come from propagating fluctuations in the accretion disk as proposed \citep{lyubarskii_flicker_1997} or modeled \citep{ingram_physical_2011}.

%\cite{stiele_energy_2015} conducted a spectral analysis of the power spectra of several black hole X-ray binaries. They observed that in the LHS power spectra described by QPO and BLN, the characteristic frequencies of BLN components in the soft energy band, exceeding 1 Hz, were consistently lower than those in the harder energy bands. In our analysis of the power spectrum of MAXI J1820$+$070, we also observed a BLN component positively correlated with changes in energy. Some past spectral analyses of MAXI J1820$+$070 (such as \citep{gao_low_2023, buisson_maxi_2019, chakraborty_spectral_2020}) in its LHS consistently indicated that the energy spectrum during the LHS is dominated by Compton components rather than thermal disk components. This suggests that the variability characteristics of the X-rays displayed in the PDS are reflective of changes in Compton photons. 

In our previous spectral analysis \citep{gao_low_2023}, we found that the broadband X-ray spectrum can be described by a thermal disc component plus a Comptonization component involving disk reflection, in which the disc component provides the seed photons for Comptonization. The results indicate that in the soft X-ray band ($<$ 10~keV), the majority of the observed photons originate from the reflection component resulting from the reprocessing of Comptonized photons by the accretion disk and direct disk photons are minimal. In the high-energy hard X-ray band ($>$ 70~keV), the spectrum is dominated by the Compton component alone. This suggests that the seed photons responsible for the $>$ 30 keV high energy Compton photons of the BLN at the lowest characteristic frequency originate from the accretion disk further out than those seed photons responsible for those $<$ 30 keV BLN photons in the processes of Comptonization and/or reflection. This also suggests that the X-ray photons producing the broadening of the characteristic frequencies of the BLNs were subjective to the origin of the seed photons and their characteristic frequencies.

\subsection{Possible interpretations}

The energy dependence of the characteristic frequencies of the BLNs shows opposite trends for the L\xb{l} and the L\xb{h} and consequent broadening of the plateau between 0.1 and 10 Hz in the PDS towards higher photon energies (see \autoref{fig:waterfall}) can be explained by the radial extension of the accretion disk corresponding to those of the seed photons and their up-scattering through an extended corona. Several recent Compton models have modeled the complex evolution of time-variable phenomena such as fractional rms amplitude and time lags spectra, especially for those of the QPOs, in black hole and neutron star X-ray binaries assuming a spherical corona geometry\citep{karpouzas_comptonizing_2020,karpouzas_variable_2021,mastichiadis_study_2022,kubota_disc_2024}, which is quite consistent with the simple geometry of corona and set up of the Comptonization we investigated here. Seed photons, which carry variability at a certain characteristic frequency of intrinsic time scales in the accretion disk will undergo a number of Compton up-scatterings on their way before they escape the corona and eventually reach the observer. The exact number of scatterings would be determined by their paths relative to the corona before traveling along the observer's line of sight, as suggested by our previous spectral investigations showed that the average optical depth was around 4.4 \citep{gao_low_2023}. Early studies of Comptonization and the angular distribution of seed photons, \cite{pozdnyakov_comptonization_1983} have shown that Comptonized photons with an optical depth exceeding four are nearly entirely emitted along the directions of the incident photons to the corona. Thus, the seed photons originating from the very innermost accretion disk can go through the corona and reach higher photon energies, generating the BLN component at the highest frequency, as in the picture suggested by \citet{stiele_energy_2015} to explain the energy dependence of the high-frequency BLN. This means that although photons in the soft and hard energy bands contribute to the same BLN component, up-scattered photons with soft energies originate from seed photons originated from the accretion disk at larger radii, and hence carry variability at lower characteristic frequencies. This explains the trend of increasing characteristic frequency with increase in the photon energy of the highest BLN as seen with \emph{Insight}-HXMT observations up to beyond 100 keV straightforwardly. 

% 第一次投稿草稿
%On the other hand, seed photons originated from the part of the accretion disk that is on the opposite side but not penetrates into the central corona can still go through the central corona, get up-scattered to high energies, and eventually run into the trajectory to the line of sight of the observer. Since these photons originate from the accretion disk further out, they only carry variability at a lower frequency and therefore can only contribute to produce the low frequency BLN. The schematic picture about the origins of the low frequency BLN and the high frequency BLN in relation to the origins of those seed photons are shown in \autoref{fig:interpretation}. To consistently explain the energy-dependent behaviours of the multiple BLNs and the energy-dependent broadening of the plateau formed by the BLNs in the PDS, we only need to acknowledge an extended disk region for providing the seed photons for up-scattering. Extending to higher photon energies, the plateau in the PDS is expected to be more broad and flat. The ultimate limits on such a broadening in terms of characteristic frequencies and total fractional rms might be set by the PLN seen in BH HSS (see \autoref{fig:interpretation}), as the PLN is formed from the thermal emission of an entire optically-thick accretion disk. In the case of MAXI~J1659$-$152 \citep{yu_energy-dependent_2013}, the PLN seen below 2 keV could have been formed by such a two-way broadening of the characteristic frequency range, with ${\rm L}_{l}$ and ${\rm L}_{h}$ going below or beyond the frequency range of interest. 

On the other hand, seed photons that originate from the accretion disk on the opposite side, which do not penetrate into the central corona, can still pass through the central corona, get up-scattered to high energies, and eventually reach the observer's line of sight. Since these photons come from regions of the accretion disk that are further out, they only carry variability at lower frequencies and therefore can only contribute to producing the low-frequency BLN. To consistently explain the energy-dependent behaviors of the multiple BLNs and the energy-dependent broadening of the plateau formed by the BLNs in the PDS, we need to acknowledge an extended disk region that provides the seed photons for the up-scattering.

%\textbf{\citet{kubota_disc_2024} constructed a passive disk model within the optically thick inner disk, which is covered by a corona that Comptonizes seed photons from the passive disk. By varying the inner disk radius and Comptonization parameters, they were able to model the spectral components of different spectral states. In our results, we interpret the energy of non-thermal X-ray photons as depending on the number of scatterings they undergo or the path length they traverse within the corona. To ensure that photons originating from disk regions not covered by the corona still undergo enough scatterings to reach higher energies, certain specific geometries of the corona, such as the vertically extended corona proposed by \citet{kara_corona_2019}, would require the corona to cover the inner accretion disk region.}

The schematic illustration of the origins of the low-frequency BLN and high-frequency BLN, in relation to the origins of these seed photons, is shown in \autoref{fig:interpretation}. As we extend to higher photon energies, the plateau in the PDS is expected to become broader and flatter. The trend of increasing frequency with increasing photon energy for the upper BLN implies potential probe of whether the accretion disk is truncated or not, and whether the actual accretion disk is truncated at the innermost stable circular orbit (ISCO) with the ${\rm L}_{h}$ determined at the highest energy band. In \citet{kara_corona_2019}, the analysis of time delays among photons in different energy bands in MAXI J1820$+$070 suggests that the accretion disk should be truncated at a smaller radius than the ISCO to account for the shorter time delay scales observed. Additional studies of the spectral properties in the LHS also support that MAXI J1820$+$070 aligns with the expectations of the truncated disk model \citep{zdziarski_accretion_2021, de_marco_inner_2021}. If the flat plateau and its frequency range in the power spectrum (power $\times$ frequency) indicate the disk range involved as we suggested, detailed characterization of the high-frequency BLN and its modeling will provide new clues to the answer of disk truncation. On the other hand, the ultimate limits on such a plateau broadening, in terms of both characteristic frequency range and the total fractional rms, might be set by the PLN observed in the BH HSS (see \autoref{fig:interpretation}), since the PLN is formed from the thermal emission of an entire optically thick accretion disk. The innermost disk radius implied by the upper BLN should correspond to the highest frequency at which the upper BLN merges under the PLN level. 

\subsection{Conclusion}
We have investigated the X-ray variability of MAXI J1820$+$070 in the LHS by modeling the PDS with three zero-centered Lorentzian components for the broad-band variability and a narrow Lorentzian component for the black hole low-frequency QPO. We found the energy-dependent behaviors of the BLN power spectral components are in fact a two-way broadening phenomenon with increasing photon energy, i.e., the L\xb{l} component shifts towards lower frequencies while the L\xb{h} component shifts towards higher frequencies. The energy-dependent frequency broadening or separation of the BLNs produces an energy-dependent extension of the power plateau in the frequency domain with increasing photon energy, indicating that the seed photons which are up-scatted to higher energies ($>$ 30 keV) originate from a wider range of the accretion disk. The energy-dependent trends of the two BLNs can be explained as due to the disk seed photons from different regions being up-scattered to the highest energies: one region corresponds to the very central inner disk embedded by the likely dense spherical corona and the other region corresponds to the disk region further out that is uncovered by the central corona; both regions of the accretion disk produce those seed photons up-scattered by the corona to the highest energies with the most number of up-scatterings. Interestingly, the trend of the two-way broadening of power spectrum would establish a power plateau in the PDS spanning over nearly two orders of magnitudes in frequency, mimic the emergence of the PLN seen in observations of MAXI~J1659$-$152 when photon energies cross the boundary $\sim$ 2 keV, which shows that the PLN was also formed by a simultaneous broadening, flattening and amplitude-decreasing of the BLNs in the power spectra towards lower photon energies (see Figure 2 and 3 in \citet{yu_energy-dependent_2013}). Such a simultaneous broadening in frequency, flattening of the power spectrum, and decreasing of fractional amplitude in the PDS might be a common process as the outcome of that the seed photons in Comptonization are from a wider, extended range in the thin disk flow. 

\section*{Acknowledgements}
We would like to thank Stefano Rappisarda, Wenda Zhang and Holger Stiele for helpful discussions. We would also like to thank the anonymous reviewer for stimulating suggestions, especially on the discussions of truncation disk models. This work made use of the data from \emph{Insight}-HXMT mission, a project funded by China National Space Administration (CNSA) and the Chinese Academy of Sciences (CAS). This work was supported in part by the Natural Science Foundation of China (grants U1838203, 12373050, 12373049 and 12361131579). 

%%%%%%%%%%%%%%%%%%%%%%%%%%%%%%%%%%%%%%%%%%%%%%%%%%
\section*{Data Availability}
The \textit{Insight}-HXMT data underlying this article are available in the public archive \url{http://archive.hxmt.cn/proposal}.

\vspace{5mm}
\facilities{HXMT}

\bibliography{cxgao_lib}
\bibliographystyle{aasjournal}

\end{document}